\documentclass{IEEEtran4PSCC}

\usepackage{verbatim}
\usepackage{cite}
\usepackage{xcolor}
\usepackage[cmex10]{amsmath}
\usepackage{bm}
\usepackage{multirow}
\usepackage{booktabs}
\usepackage{caption}
\usepackage{subcaption}
\usepackage{soul}

\ifCLASSINFOpdf
   \usepackage[pdftex]{graphicx}
\else
   \usepackage[dvips]{graphicx}
\fi

\usepackage{url}

\makeatletter
\let\old@ps@headings\ps@headings
\let\old@ps@IEEEtitlepagestyle\ps@IEEEtitlepagestyle
\def\psccfooter#1{%
    \def\ps@headings{%
        \old@ps@headings%
        \def\@oddfoot{\strut\hfill#1\hfill\strut}%
        \def\@evenfoot{\strut\hfill#1\hfill\strut}%
    }%
    \def\ps@IEEEtitlepagestyle{%
        \old@ps@IEEEtitlepagestyle%
        \def\@oddfoot{\strut\hfill#1\hfill\strut}%
        \def\@evenfoot{\strut\hfill#1\hfill\strut}%
    }%
    \ps@headings%
}
\makeatother

\psccfooter{%
        \parbox{\textwidth}{\hrulefill \\ \small{22nd Power Systems Computation Conference} \hfill \begin{minipage}{0.2\textwidth}\centering \vspace*{4pt} \includegraphics[scale=0.06]{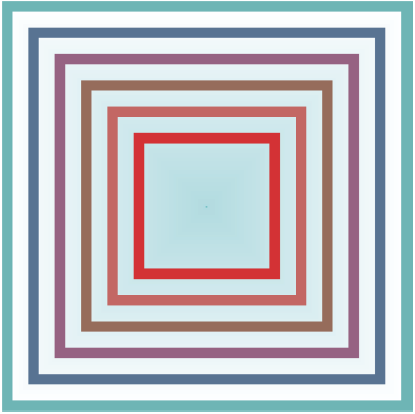}\\\small{PSCC 2022} \end{minipage} \hfill \small{Porto, Portugal --- June 27 -- July 1, 2022}}%
}

\begin{document}

\title{A Framework for Risk Assessment and Optimal Line Upgrade Selection to Mitigate Wildfire Risk}

\author{
\IEEEauthorblockN{Sofia Taylor and Line A. Roald}
\IEEEauthorblockA{Department of Electrical and Computer Engineering\\
University of Wisconsin-Madison,
United States\\
\{smtaylor8, roald\}@wisc.edu}
}

\maketitle

\begin{abstract}
As wildfires in the United States are becoming more frequent and severe, mitigating the risk of wildfire ignitions from power line faults is an increasingly crucial effort. Long-term ignition prevention strategies, especially converting overhead lines to underground cables, are expensive. Thus, it is important to prioritize upgrades on lines that will reduce wildfire ignition risk the most. However, since so many factors contribute to ignition risk, it is difficult to quantify the wildfire risk associated with power lines. This paper examines how various risk definitions based on historical wildfire risk maps can be used to inform transmission upgrade planning. These risk metrics are evaluated using an optimization model that determines which overhead lines should be undergrounded such that the total wildfire risk in the network is minimized. The risk assignment and upgrade selection are tested on both a synthetic network and the actual transmission lines in California.
\end{abstract}


\thanksto{\noindent 
Submitted to the 22nd Power Systems Computation Conference (PSCC 2022).\\
This work is funded by the National Science Foundation (NSF) under Grant. No. ECCS-2045860
and partially supported by the NSF Graduate Research Fellowship Program under Grant No. DGE-1747503. Any opinions, findings, and conclusions or recommendations expressed in this material are those of the authors and do not necessarily reflect the views of the NSF.}

\section{Introduction}
In the United States, the total area burned by wildfires, wildfire frequency, and federal fire suppression costs per year have increased significantly since 1985 \cite{NIFC}. Wildfire prevention is an increasingly crucial effort, especially as climate change exacerbates future fire risk conditions \cite{Martinuzzi2019future}. 
Power line faults are one of the major sources of wildfire ignitions \cite{syphard2015location}. Downed lines, vegetation contact, conductor slap, or component failures can produce %
fault currents and sparks that 
may ignite fires under hot, dry, and windy conditions\cite{russell2012distribution, jazebi2020}. The deadliest and most destructive wildfire in California's history, the 2018 Camp Fire, was ignited by an aging transmission line \cite{CalFire}. 

A particular challenge of ignitions from electric power lines is that a common factor -- high wind speeds -- increases both the probability of ignitions due to electric faults \emph{and} promotes a rapid spread of the resulting fire. As a result, wildfires ignited by power lines tend to be larger than fires from other causes \cite{miller2017electrically}. For example, wildfires ignited by power lines in San Diego County account for only 5\% of all ignitions, but 25\% of the total acres burned \cite{syphard2015location}.

\begin{figure}
    \centering
    \includegraphics[width=\columnwidth]{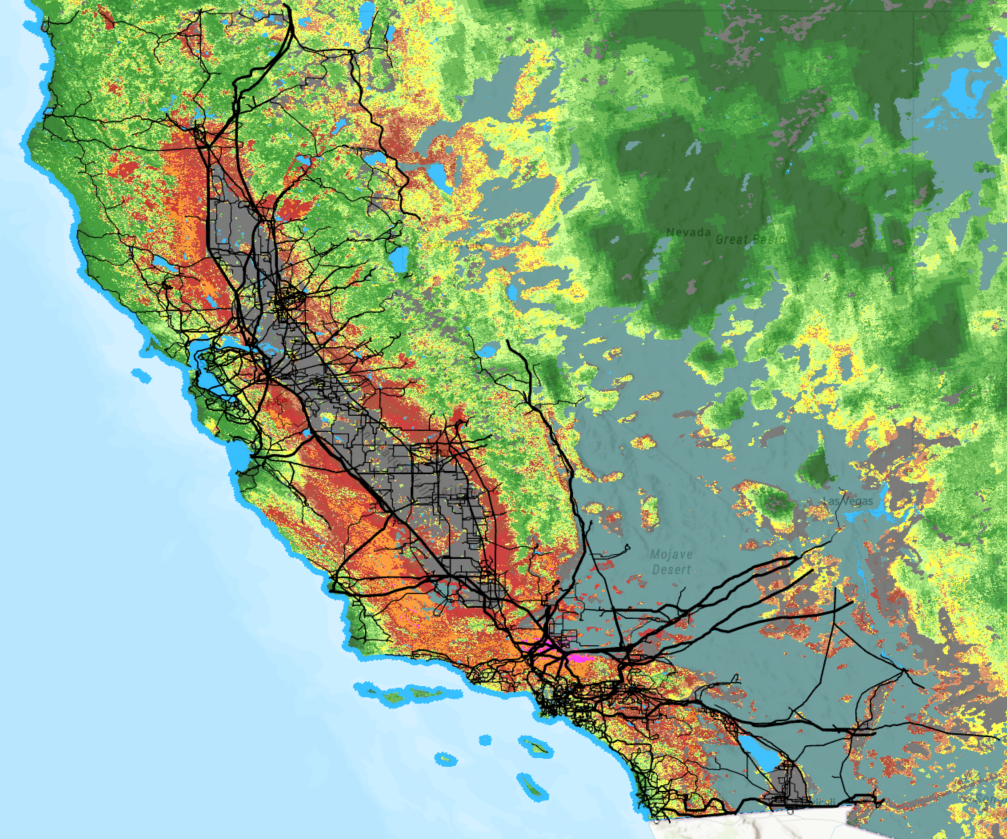}
    \caption{\small Overlay of California's transmission lines \cite{CECdata} with the Wildland Fire Potential Index map for August 1st, 2021 \cite{WFPI}. Warmer colors indicate higher wildfire potential.} 
    \label{fig:WFPI_with_CEC}
\end{figure}

Several strategies may prevent ignitions from electric infrastructure \cite{jazebi2020}. %
During high risk conditions, utilities currently implement public safety power shutoffs \cite{PSPS}, which de-energize the lines in high risk zones to avoid the release of fault currents and prevent wildfire ignitions. While preemptive shutoffs are effective in preventing ignitions, they can result in wide-spread power outages \cite{CPUCpspsrollup}. %
This consequence is particularly harmful for people that depend on electric medical devices and members of socially vulnerable communities \cite{IEJ2020}, and still leaves the population exposed to non-power line ignitions. 
Results in \cite{rhodes2020balancing} showed that it is possible to reduce both wildfire risk and the size of power outages by incorporating power flow modeling in shutoff decisions.  %

Less disruptive ignition prevention strategies include vegetation management, replacing aging components, and converting overhead power lines to underground cables \cite{pg&e2021wildfire, PGE-undergrounding}. 
Undergrounding is attractive because it reduces the need for costly short- and long-term ignition prevention strategies in the future,
as the ignition risk is essentially reduced to zero once the line is undergrounded. %
Further, underground cables are less susceptible to impacts from wildfires (e.g., flashovers due to air pollution \cite{zhang2011statistical} or fire damage to towers \cite{sathaye2013rising}), offering another argument for undergrounding in areas with high wildfire exposure.
Thus, undergrounding is seen as a highly effective, though expensive, measure to reduce mitigate power line-wildfire interactions. 

However, since undergrounding the entire electric grid is prohibitively expensive,     %
we need to select lines that exhibit the highest risk. 
Assessing the long-term ignition risk associated with a particular power line is challenging due to the complex and time-varying nature of wildfire risk. 
Following the 2009 Black Saturday bushfires in Victoria, Australia, a database of historical weather, fault, and terrain information was compiled and used, along with accurate grid models and fire behavior simulation, to quantify the ignition risk reduction associated with hardening strategies \cite{Huston2015assessing}. Compiling and processing such data from multiple entities is a time-consuming and challenging task,
and requires resources and know-how that most utilities may not have access to.
Further, the analysis in \cite{Huston2015assessing} is very specific to multiple sources of regional data, making it difficult to apply to other locations. In the United States, there are currently no publicly available databases specifically for quantifying the ignition risk of power lines and hardening strategies.
Thus, there is a need among grid planners for more accessible and flexible methods of defining wildfire risk in the context of upgrade selection.

This paper addresses this gap by proposing a framework for assessing the wildfire risk associated with power lines, as well as an optimization problem to select an optimal set of lines for upgrading. The framework defines wildfire risk as a function of two components: the probability of electric faults leading to ignitions, and the potential for large wildfires and fire spread in the area around the line. The probability of electric faults leading to ignitions is hard to assess for individual lines without detailed information typically available only to utilities. However, we can incorporate known trends. For example, per mile of power line, distribution lines are three times more likely to cause ignitions compared with transmission lines \cite{PGE-amendedsafetyplan}.  
While electric fault and ignition probabilities are hard to assess, data on the potential for large fires and fire spread is readily available. In particular, the Wildland Fire Potential Index (WFPI) is a publicly available metric that quantifies current fire risk based on satellite imagery and local vegetation and weather conditions \cite{WFPI}. 
The WFPI is published daily by the United States Geological Survey (USGS) as a map covering the contiguous United States. It is one of several wildfire risk databases and maps that model general (not power-specific) wildfire risk. 
Using geographic information system (GIS) software, such data can be leveraged to assess the wildfire risk in the area around power lines.

Even with access to this spatio-temporal data on ignition probabilities and wildfire risk, important questions about how to quantify wildfire risk remain. Given the geographical extent of a transmission line, there are several ways in which we can translate risk along the line into a single risk metric, reflecting different approaches to measure and mitigate risk. For example, limiting the probability of ignitions from lines that experience (possibly infrequent) episodes of extreme wildfire risk will lead to different results compared with limiting the probability of ignitions for lines that generally have a continuously high, but never extreme wildfire risk. Although the choice of the risk metric may have significant impact on how risk mitigation strategies such as undergrounding are implemented, we are not aware of existing work that discusses different metrics or investigated their impact on a decision making process.

In this paper, we take an initial step towards filling this gap. The first contribution of our paper is to examine multiple methods of defining the wildfire ignition risk associated with power lines. As a second contribution, we incorporate these metrics as inputs to an optimization model which identifies the optimal set of power lines to underground to reduce wildfire risk, while not exceeding a pre-defined budget. %
The model considers multiple wildfire risk scenarios to determine the line upgrades that provide the best improvements across all of the scenarios.

Finally, we demonstrate our method through a case study based on the RTS-GMLC grid \cite{RTSGMLC} and the real transmission lines in California \cite{CECdata}, in which we analyze different risk metrics and assess how sensitive the upgrade selection is to various model parameters. The data files produced in this study, which contain the risk values assigned to power lines across multiple scenarios, are available for public use \cite{repo}.
In summary, the contribution of this paper is (1) a discussion on how to define and compute metrics to quantify the risk of wildfire ignitions from power lines, and (2) a simple, yet effective and flexible method to plan undergrounding of power lines in wildfire prone areas.

The rest of this paper is organized as follows. Section \ref{sec:formulation} details the optimization model formulation, while Section \ref{sec:risk} describes how to assign risk values to power lines. Section \ref{sec:processing} details the data, software, and processing steps used for risk assignment, while Section \ref{sec:results} presents results obtained by applying the proposed risk definitions and upgrade selection method. Section \ref{sec:conclusion} summarizes and concludes. %

\section{Identifying the Optimal Set of Power Lines for Undergrounding}
\label{sec:formulation}

In this section, we present a simple optimization-based method to demonstrate how data on wildfire risk can be used to support decisions regarding \emph{which power lines} should be prioritized for undergrounding in a region with significant wildfire risk. %
We consider a power system with $N$ overhead power lines. In most cases, a utility likely would not underground the entire length of a long transmission line due to high cost. Therefore, we split the lines up into line segments represented by the set $\mathcal{L}$. Note that the number of line segments $N_L = |\mathcal{L}| \geq N$.
For each line segment $l\in\mathcal{L}$, we define the associated wildfire risk as $R_{l}$ and the cost of undergrounding by $C_l$. 
Further, we define a binary decision variable $z_l$ which represents whether line segment $l$ is undergrounded ($z_l=1$) or not ($z_l=0$). 
We represent the budget for undergrounding lines as $C^{\max}$.

With this information, we formulate a simple optimization problem to identify the optimal set of lines to underground:
\begin{subequations}\label{upgradeopt}
\begin{align}
    \min_{z_l} ~&
    \rho(R_l,z_l)
    \label{eq:obj} \\
    \text{s.t. }~ &\sum_{l=1}^{N_L} C_l z_l \leq C^{\max} \label{eq:budget} \\
    & z_l \in \{0,1\} \quad \forall l\in\mathcal{L}. \label{eq:int}
\end{align}
\end{subequations}
Here, the objective function \eqref{eq:obj} minimizes the total wildfire risk in the system, represented by the function $\rho(R_l, z_l)$, which depends on the wildfire risk $R_l$ and the upgrade status $z_l$ of each line.
The constraint \eqref{eq:budget} limits the upgrades to those that are possible within the defined budget, while \eqref{eq:int} requires the undergrounding variables $z_l$ for line segment $l$ to take on values of either 0 or 1 (i.e., partial undergrounding is not possible). Assuming that $\rho(R_l, z_l)$ can be expressed as a linear function, this simple model is a version of the classical 0-1 knapsack problem.

In this model, the risk values $R_l$ of a transmission line $l$ represent a single, aggregate measure of risk across the entire geographical span of the line and across multiple scenarios of daily wildfire risk. The wildfire risk function $\rho(R_l, z_l)$ further aggregates the risk values for all the individual lines into a single value. The method used to aggregate wildfire risk metrics can have a significant impact on the results. 
In the next section, we discuss different options for defining $R_l$ and $\rho(R_l, z_l)$ in detail.

\section{Assessing Wildfire Risk of Transmission Lines}
\label{sec:risk}

Obtaining and synthesizing information about all of the factors that impact wildfire risk from electric grids and deriving an aggregate risk value for the entire length of the line is a challenging and time-consuming task. Furthermore, since decisions of which lines should be put underground is a long-term planning problem, we need to consider how wildfire risk varies over time. 
To address this challenge, we divide task of defining and computing $\rho(R_l, z_l)$ in three parts. First, we discuss how to compute risk for a small line segment at a given point in time. Then, we discuss how these risk values can be aggregated geographically (i.e., along the length of the line) and in time (i.e., across multiple scenarios) for a single line. Finally, we discuss how to aggregate risk across all lines in the network to define the total system risk $\rho(R_l, z_l)$.

\subsection{Wildfire risk for a small line fragment}
Risk modeling in general considers both the \emph{probability} and the \emph{impact} of an event. 
In our context, this corresponds to the probability of an ignition, which we denote as $\pi_{i,s}$ for a small line fragment $d\ell$, and the impact of the resulting fire at this location, denoted by the wildfire potential $w_{i,s}$. The subscript $s$ refers to these values evaluated for a specific day (or scenario) $s$. We define the corresponding wildfire risk $r_{i,s}$ for the line fragment $i$ as the product of these two terms, i.e.
\begin{equation}
    r_{i,s}= \pi_{i,s} \cdot w_{s,i}
\label{risk1}
\end{equation}
We note that the above definition is more comprehensive than adapting a definition that aims at, e.g., minimizing the frequency of ignitions (since this metric would not capture the difference in impact of ignitions at different times and locations).
We next discuss the meaning of $w_{i,s}$ and $\pi_{i,s}$ in more detail.

\subsubsection{Wildfire potential $w_{i,s}$}
There are several publicly available resources for wildfire risk, such as \cite{WHP, WFPI, Texas-WRAP}, that quantify the risk of ignition and fire spread from any source (not specific to power lines) based on natural risk factors.
Here, we choose to work with the WFPI maps produced by the USGS \cite{WFPI}. These maps are published once daily for the contiguous United States and include both current and 7 days into the future, as well as historical records of wildfire data.
The WFPI maps represent fire potential in terms of unit-less values from 0 to 150, with higher values representing higher potential. The indices are computed using an algorithm with variables for fuel conditions, rainfall, temperature, and wind speed, which are quantified using satellite observation, fuel models, and weather forecasts. There are also indices from 248 to 254 that represent land classified as `Cloud', `Outside US', `Barren', `Ag Land', `Marsh', and `Water'. 
Fig. \ref{fig:WFPI_with_CEC} shows the WFPI map for August 1, 2021 underneath a map of California's transmission lines from the California Energy Commission (CEC) \cite{CECdata}. 
By overlaying the WFPI map for a specific day $s$ and electric grid models via GIS software, we can obtain an estimation of the WFPI-based wildfire potential associated with the power line fragment $w_{i,s}$. The steps needed to obtain these values are discussed in detail in Sec. \ref{sec:processing}.

There are several important advantages of using the WFPI data instead of other wildfire risk sources. First, WFPI does not account for historical data on wildfire ignitions. Other indices that do account for historical ignitions typically consider all ignitions, including ignitions from sources like lightning or arson, which are not related to electric power grids. 
Second, although the WFPI does not account for ignitions, validation against historical data has shown that higher percentages of large fires and fire spread occurred at the higher WFPI values \cite{WFPI}, demonstrating that WFPI is an effective metric of wildfire potential.
Third, the WFPI maps cover the contiguous United States. Thus, our methodology for assignment of risk values $w_{i,s}$ can be directly adopted by utilities anywhere in the United States.

\subsubsection{Probability of ignitions $\pi_{i,s}$}
The main avenue to reduce risk is to reduce the probability of electric faults, which may lead to wildfire ignitions.
Electric utilities control several factors that impact the prevalence of power line faults, including the line voltage, width of the right-of-way, component age and function, state of vegetation management and type of protection equipment. The management of these factors have received significant attention in the context of wildfire risk mitigation \cite{PGE-amendedsafetyplan, jazebi2020}.
Recent research has made significant progress in predicting outages \cite{dokic2019spatially}, which could be a useful step towards predicting the time-varying probability of ignitions $\pi_{i,s}$.
However, predicting outages is not the same as estimating the probability of faults (as not all faults lead to an outage) or the probability of ignitions $\pi_{i,s}$ (as not all faults lead to an ignition). Furthermore, these methods require access to internal data such as vegetation management, maintenance status, or historical outage records. An accurate determination of $\pi_{i,s}$ is thus out of scope for this paper.

Fortunately, when making decisions about which lines to upgrade, the \emph{relative} risk and relative ignition probabilities of different line segments is more important than the absolute value. Therefore, we can leverage aggregate statistics of wildfire ignitions from electric power infrastructure published by utilities to obtain estimates of the relative ignition probabilities of different groups of lines. Specifically, publicly available documents released by PG\&E indicate that distribution lines are three times more likely to cause wildfire ignitions per mile of line, as compared with transmission lines\footnote{This is likely due to the relatively short distance between distribution conductors and surrounding vegetation, and larger number of components that could fail (more towers, insulators, etc.).} \cite{PGE-amendedsafetyplan}. If we assume that all transmission line fragments have a common base probability $\pi$ of causing ignitions, we can express the relative probability of an ignition for individual line fragments based on their rated voltage level in kV,
\begin{equation}
    {\pi_{i,s}}= \begin{cases}
            \pi,& \text{if voltage level } \geq V_{dist}^{\max} \\
            3\cdot \pi, & \text{if voltage level } < V_{dist}^{\max},
            \end{cases} \label{eq:voltage}
\end{equation}
where $V_{dist}^{\max}$ represents the highest distribution system voltage.

\subsection{Aggregating wildfire risk along a line}
We next consider the task of aggregating the wildfire risk along a longer line segment. 
Considering a WFPI map for a particular scenario $s$, we define the wildfire risk $r_{l,s}$ for a given line segment $l$ and scenario $s$ by combining the geographical information from the WFPI maps with the geographical path of the line. 
We denote the set of small line fragments $d\ell$ that make up the longer line segment $l$ by $i\in\mathcal{I}_l$. For each line fragment $d\ell$, we assume that we have access to a wildfire risk $r_{i,s}$, defined according to \eqref{risk1}. 
We propose two different ways of aggregating the risk along the line:
\begin{itemize}
    \item \emph{Maximum risk} defines the wildfire risk $r_{l,s}$ as the maximum wildfire potential value along the line, i.e., 
    \begin{equation}
        r_{l,s}^{\max} = \max_{i\in\mathcal{I}_l} \{r_{i,s}\}.
    \end{equation}
    This risk definition is useful if we want to target our mitigation efforts on lines that experience extreme wildfire risk somewhere along the line.
    
    \item \emph{Cumulative risk metric} defines the wildfire risk $r_{l,s}$ as the integral of the wildfire risk $r_{i,s}$ along the line segment from the line segment start $\ell_0$ to the line segment end $\ell_f$. This integral is calculated as the
    sum %
    of the wildfire risk $r_i$ of all line fragments belonging to the segment $l$, 
    \begin{equation}
        r_{l,s}^{\text{cum}} = \int_{\ell_0}^{\ell_f} r_{i,s} \,d\ell = \sum_{i\in\mathcal{I}_l} r_{i,s}. \label{eq:cum}
    \end{equation}
    This risk metric aggregates the wildfire risk assuming that the risk values are additive. For example, two line fragments $1,2$ with risk values $r_1=r_2=70$ represent the same risk as two line segments with risk values $r_1=140$ and $r_2=0$. 
    We note that this risk value will tend to be longer for longer line segments. However, we do not normalize the risk by the length of the line segment, as the segment length is considered in the budget constraint \eqref{eq:budget} where a longer segment typically will cost more to underground.
    
\end{itemize}

We note that it is possible to define other aggregated risk metrics $r_{l,s}$. For example, we could define a conditional version of the cumulative risk metric that only considers risk values as non-zero if they are greater than a threshold $r_i^{\min}$. We leave a detailed discussion of these additional metrics for future work, and restrict our investigations to the maximum and cumulative metrics for risk aggregation along the line.

\subsection{Aggregating wildfire risk across multiple scenarios}
From the above section, we obtain a wildfire risk $r_{l,s}$ for a specific line segment $l$ and a scenario $s$. 
Next, we consider how to aggregate this risk in time (i.e., across multiple scenarios) to obtain a single risk value $R_l$ for each line segment $l$.
Similar to the aggregation along the line segment, we consider two risk metrics, namely the worst-case risk and the cumulative risk.
\begin{itemize}
    \item \emph{Worst-case risk} is defined as the maximum risk across all line segments and scenarios, 
    \begin{equation}
        R_{l}^{\max} = \underset{s\in\mathcal{S}}\max\{ r_{l,s}^{\max}\}.
        \label{eq:risk-line-max}
    \end{equation}
    \item \emph{Cumulative risk} is approximated as the sum (or cumulative) risk across all scenarios,
    \begin{equation}
        R_{l}^{\text{cum}} = \underset{s\in\mathcal{S}}\sum r_{l,s}^{\text{cum}}.
        \label{eq:risk-line-cum}
    \end{equation}
\end{itemize}

\subsection{Computing total system risk}
Given the worst-case risk $R_{l}^{\max}$ and cumulative risk $R_{l}^{\text{cum}}$ for a given line segment $l$, we calculate the worst-case and cumulative risk for the total system. These metrics are calculating with and without consideration of undergrounding of lines.
\begin{itemize}
    \item \emph{Worst-case risk:} The worst-case risk for the total system \emph{without} undergrounding of lines is given by 
    \begin{equation}
        R^{\max}_{\text{WO}} = \underset{l\in\mathcal{L}}\max\{ R_{l}^{\max}\} %
    \end{equation}
    which is equivalent to the maximum risk value occurring across all line fragments and all scenarios. The worst-case risk for the total system \emph{with} undergrounding of lines is given by 
    \begin{equation}
        R^{\max}_{\text{W}} = \underset{l\in\mathcal{L}}\max\{ (1-z_l) R_{l}^{\max}\}.
        \label{eq:MMM}
    \end{equation}
    Here, if line segment $l$ is chosen for undergrounding, $z_l=1$ and wildfire risk for this segment is zero.
    \item \emph{Cumulative risk:} The cumulative total system risk \emph{without} undergrounding of lines is approximated by
    \begin{equation}
        R^{\text{cum}}_{\text{WO}} = \underset{l\in\mathcal{L}}\sum R_{l}^{\text{cum}} 
    \end{equation}
    which represents the cumulative risk summed across all lines. The cumulative total system risk \emph{with} undergrounding of lines is approximated by
    \begin{equation}
        R^{\text{cum}}_{\text{W}} = \underset{l\in\mathcal{L}}\sum (1-z_l) R_{l}^{\text{cum}}.
        \label{eq:CCC}
    \end{equation}
\end{itemize}

Based on the above definitions of risk, we define the objective function $\rho(R_l, z_l)$ as a the convex combination of normalized versions of the two risk metrics, i.e.,
\begin{align}
    \rho(R_l, z_l) &= (1-\alpha) \frac{R^{\text{cum}}_{\text{W}}}{R^{\text{cum}}_{\text{WO}}} + \alpha \frac{R^{\max}_{\text{W}}}{R^{\max}_{\text{WO}}} \nonumber %
\end{align}
Here, $\alpha$ is a trade-off parameter that represents whether we want to focus on minimizing the cumulative risk ($\alpha \rightarrow 0$) or the maximum risk ($\alpha \rightarrow 1$).
The benefit of using normalization is that both $R^{\text{cum}}_{\text{W}}/R^{\text{cum}}_{\text{WO}}$ and $R^{\max}_{\text{W}}/R^{\max}_{\text{WO}}$ are normalized to values between $0$ and $1$. Furthermore, the normalization cancels out the value of the transmission line ignition probability $\pi$.
We note that in our implementation, to avoid numerical issues in the solver, we scaled the objective by multiplying with $R^{\text{cum}}_{\text{WO}}$.

\subsection{Problem variants}
Based on how the parameter $\alpha$ is chosen in the objective function, we define three different versions of the problem \eqref{upgradeopt}:
\begin{itemize}
    \item \emph{Max-max-max (MMM):} When $\alpha=1$ we consider only the maximum wildfire risk. We refer to this problem variant as Max-Max-Max (MMM) because it takes the maximum value across all line fragments, all scenarios and all lines. The corresponding MMM risk metric is the risk as defined by \eqref{eq:MMM}.
    \item \emph{Cumulative-cumulative-cumulative (CCC):} When $\alpha=0$ we consider only the cumulative wildfire risk. We refer to this problem variant as Cumulative-Cumulative-Cumulative (CCC) because it considers the cumulative risk across all line segments, lines and scenarios. The corresponding CCC risk metric is defined by \eqref{eq:CCC}.
    \item \emph{Trade-Off:} If we pick an intermediate $0<\alpha<1$, we obtain a solution that represents a trade-off between the two metrics above. We therefore refer to this problem variant as the trade-off variant. 
\end{itemize}
In the case study, we assess how the different problem variants, corresponding to different choices of $\alpha$, lead to different upgrade decisions.

\section{Data Processing and Risk Assignment}
\label{sec:processing}
The risk assignment methods and upgrade selection model are demonstrated using WFPI maps and two transmission line networks. This section details the data, software, and processing steps needed to assign risk to line segments. The data files produced in this study are available for public use \cite{repo}.

\subsubsection{Data and Software} 
Two months of WFPI data (for July and August 2021) are used as test case inputs. The proposed methods are demonstrated on two systems in the western US. The first is the RTS-GMLC, a 72-bus system with an artificial location in southern California, Nevada, and Arizona \cite{RTSGMLC}. The second system is the actual grid in California, where the location of transmission lines have been 
made publicly available through the CEC \cite{CECdata}. Both systems were selected for their GIS-compatible file formats and locations in fire-prone regions.

The software application ArcGIS Pro is used to process data and assign risk values to the power lines in the test cases. The steps for this processing and assignment, using ArcGIS Pro's built-in geoprocessing tools and Python programming interface, are described below.

\subsubsection{Pre-processing of WFPI data}
The USGS publishes the WFPI maps in the TIF format, which can be added to an ArcGIS Pro project by creating a new raster layer with the function \verb+MakeRasterLayer+.
To further work with this data, we first convert the file from a raster layer to a feature class using the function \verb+RasterToPolygon+.
There are several high-value risk indices in the WFPI data that indicate land with very low or no wildfire risk. We replace the values of these indices with zeros to indicate non-existing risk and simplify the following calculations. %
This is accomplished with the \verb+UpdateCursor+ iterator.

\subsubsection{Importing power line data}
Next,  we add the grid files for RTS-GMLC and the CEC lines to the current map in ArcGIS Pro. The RTS-GMLC is published in the GeoJSON format and must be converted to a feature class with the \verb+JSONToFeatures+ function. 
The CEC lines are added directly as a shape file. To facilitate risk value assignment, the electric grid data should be stored in a polyline feature class. 

\subsubsection{Generating line segments}
Next, we segment the power lines in both grids. To do this, we first use \verb+GeneratePointsAlongLines+ to generate points along the lines 
at the desired interval (1 and 10 km in this case). Then, we split the lines at those points and store the segments in a new feature class using \verb+SplitLineAtPoint+.

\subsubsection{Assigning risk values}
Finally, we assign risk values to the power lines or power line segments remains. 
We first compute $r_{l,s}$ using the maximum risk metric and the cumulative risk metric for each line and each scenario.

For the maximum metric, spatially join the line features with each WFPI map, or scenario $s\in\mathcal{S}$. %
This can be accomplished with the \verb+SpatialJoin+ function
with the merge rule of ``max" for the WFPI gridcode. 

For the cumulative metric, there is no simple merge rule and we have to compute the metric in several steps. First, we apply the \verb+TabulateIntersection+ tool for the line features and each WFPI map, creating a new table for each scenario $s\in\mathcal{S}$. Then, in each table, we multiply every WFPI gridcode by the corresponding length. Finally, we compute the sum of these products for each line segment ID to obtain $r_{l,s}$. %

\section{Test Case Results}
\label{sec:results}
The proposed risk assignment and upgrade selection methods are demonstrated through several test cases. First, a simple example using the CCC problem variant is presented. This is then expanded to show the effect of using different line segment lengths in the problem. Next, the performance of the CCC problem is compared with the MMM problem, along with the tradeoff between the two. Finally, an example that weighs risk values based on line voltage is presented to demonstrate how grid-specific data can be incorporated to quantify the relative probability of electric fault.

\subsection{Simple Example}
In this initial example, we solve the CCC problem variant for the RTS-GMLC system with non-segmented lines.
Daily WFPI maps from the full 2021 year are used. Note that we assign risk values of zero to the transformers, which are included in the RTS branch dataset as lines of zero-length. %
For simplicity, the upgrade cost for each line is assumed to be 2 million USD per mile, in accordance with PG\&E estimates from \cite{2mil}. The total length of lines that can be upgraded is bound by a budget, which we choose here to be 600 million USD.

\begin{figure}
    \centering
    \includegraphics[width=\columnwidth]{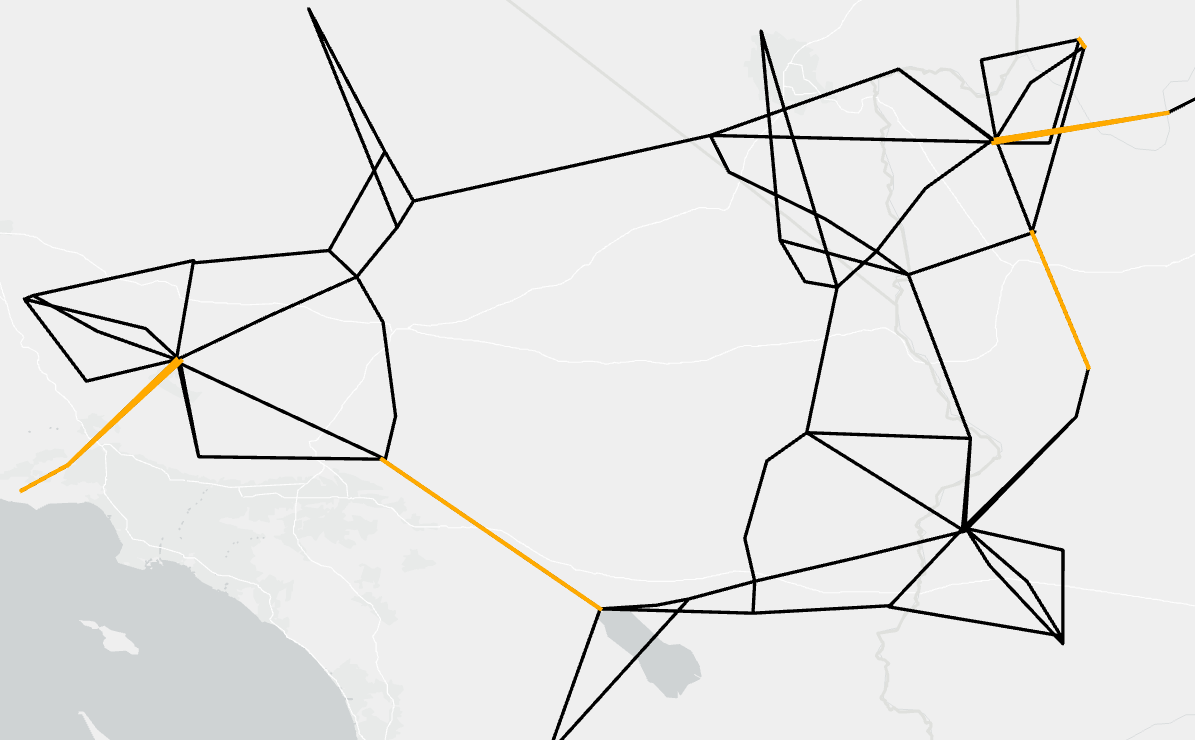}
    \caption{\small Line segments selected for upgrade (highlighted in orange) in simple example.}
    \label{fig:ex_map}
    \vspace{-10pt}
\end{figure}

In this example, the CCC model chooses 8 lines to be upgraded out of the total 104, as shown in Fig. \ref{fig:ex_map}. This corresponds to 300 miles, or 9\% of the total line length.
Among the 8 lines chosen for undergrounding, the average cumulative risk value $R_{l}^{\text{cum}}$ is 4.70e6. The standard deviation of $r_{l,s}^{\text{cum}}$ across different scenarios is 1.45e6. 
Among the remaining 96 lines, the average cumulative risk value $R_{l}^{\text{cum}}$ is 1.06e6, and $r_{l,s}^{\text{cum}}$ has a standard deviation of 0.305e6 across scenarios. 
From these numbers, we conclude that our method successfully chooses high risk lines for undergrounding. We also see that the wildfire risk variability across different days is higher for the high risk lines (i.e., they have a larger standard deviation across days), which is as expected since there is not always high wildfire risk conditions.

\subsection{Segmenting lines}
In the previous example, we considered undergrounding decisions for entire transmission lines. However, we could also consider shorter line segments by splitting the lines into pieces with a length of at most 10km or 1km.
Using shorter line segments allows us to develop a more detailed plan and more carefully target undergrounding efforts in the highest risk zones. However, shorter line segments increase the computational requirements of our model, both in data processing (i.e., obtaining risk values $R_l^{\text{cum}}$ and $R_l^{\text{max}}$ values for a large number of line segments $l$) and in the optimization problem (i.e. each line segment requires the introduction of a binary decision variable $z_l$ to represent whether or not that segment should be undergrounded).

\begin{figure}
    \centering
    \includegraphics[width=\columnwidth]{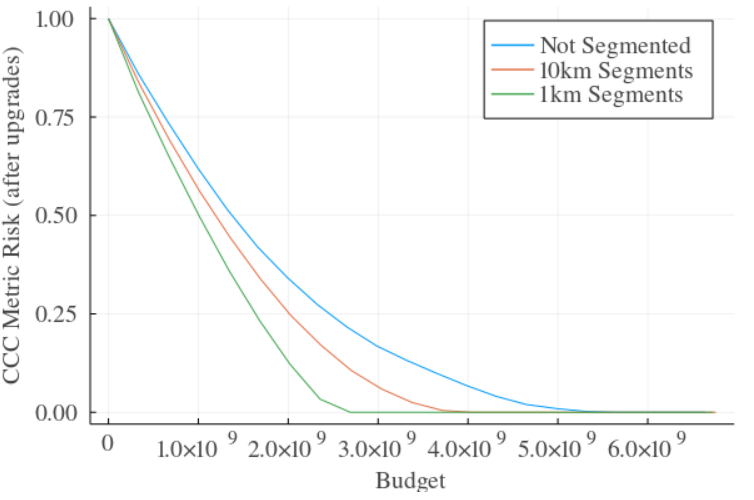}
    \caption{\small Plot of the normalized system-wide risk (as defined by the CCC problem variant) after upgrades versus cost, with three series for no line segmentation, 10-km segments, and 1-km segments.}
    \label{fig:compare_segments}
\end{figure}

Given these trade-offs, we study the effect of line segment length on upgrade selection %
in the RTS-GMLC, using the CCC problem variant.
Fig. \ref{fig:compare_segments} shows the post-upgrade risk for budgets ranging from 0 to 6.7 billion USD (the total cost of upgrading all lines) for three cases: the full lines, 10-km line segments, and 1-km line segments. 
From the plot, we can see that assigning risk to and choosing upgrades from a collection of shorter line segments allows for greater risk reduction for a given budget. 

\subsection{Spatio-Temporal Variability in the MMM and CCC risk metrics}
\begin{figure*}
    \begin{subfigure}[b]{0.5\textwidth}
        \centering
        \includegraphics[width=\columnwidth]{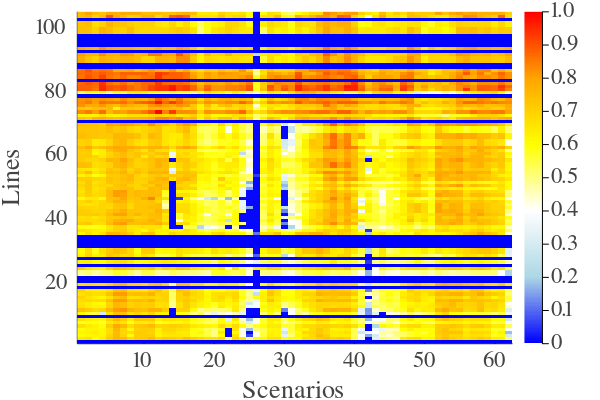}
        \caption{\small Maximum risk (normalized)}
        \label{fig:heat_max}
    \end{subfigure}%
    \hfill
    \begin{subfigure}[b]{0.5\textwidth}
        \centering
        \includegraphics[width=\columnwidth]{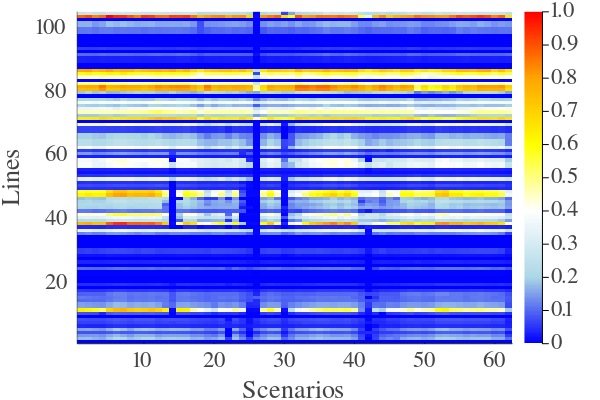}
        \caption{\small Cumulative risk (normalized)}
        \label{fig:heat_cm}
    \end{subfigure}%
    \hfill
    \hfill
        \caption{\small Heat maps of the normalized risk values assigned to non-segmented RTS-GMLC lines, based on the  (a) maximum risk definition and (b) cumulative risk definition along each line for the WFPI maps in July and August 2021. Warmer colors indicate higher risk.}
        \label{fig:Heat_maps}
\end{figure*}

WFPI maps, which exhibit temporal and spatial variability, are used to assign risk to the lines in the RTS-GMLC synthetic grid. To visualize how risk values vary across time and across non-segmented lines in the system, heat maps are produced for the cumulative and maximum risk along each line. These heat maps, which include just two months (i.e., 60 scenarios where each scenario corresponds to one day) for visibility, are in Fig. \ref{fig:Heat_maps}.
In both heat maps, it is clear that there are several days (represented as scenarios along the x-axis) where most of the lines have relatively low CCC and MMM risk values, possibly due to rainy, humid, cool, and low-wind conditions. This temporal variation demonstrates the importance of examining many days of risk maps so that both low and high risk days are represented in the data. When comparing differences in risk values along the y-axis, we observe that certain lines that have relatively high risk and others that have relatively low risk across all the full period. %
Another observation is that the MMM risk metric results in less variation between different lines compared to the CCC metric. %

\subsection{Comparing solutions with varying trade-off parameter $\alpha$}
\begin{figure}
    \centering
    \includegraphics[width=0.9\columnwidth]{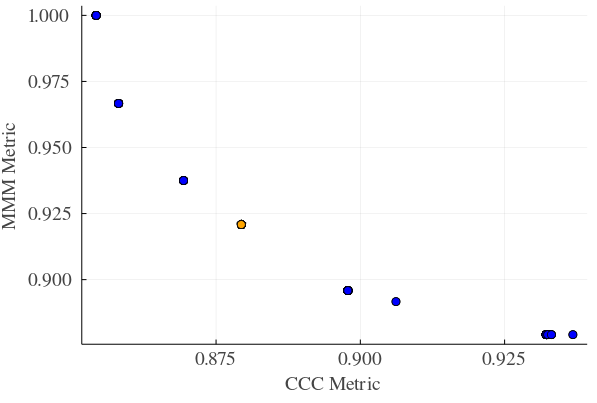}
    \caption{\small Pareto curve of normalized system-wide risk values that result from different values of the trade-off parameter. The orange point corresponds to $\alpha = 0.4$, which is the case used for comparison with the other risk metrics.}
    \label{fig:tradeoff_plot}
\end{figure}

Next, we examine the trade-off betwewen the CCC and MMM risk metrics by varying the trade-off parameter $\alpha$ between 0 and 1.
Fig. \ref{fig:tradeoff_plot} shows a Pareto plot the solutions that result from varying $\alpha$. Each point on the curve represents a solution where lines are selected for upgrade such that both the CCC and MMM risk are minimized, however, the relative importance of each metric is changing as $\alpha$ changes, leading to different solutions. %

To examine the solutions in more detail, we solve the problem once for each variant assuming a constant budget of 300 million USD. For the trade-off formulation, the trade-off parameter of $\alpha = 0.4$, which corresponds to the orange point in Fig \ref{fig:tradeoff_plot}. It was chosen for as a solution that shows significant (although not exactly equal) reductions in both the CCC- and MMM-based risk.
In each case, the reductions of maximum risk and cumulative risk are computed based on the selected upgrades and are summarized in Table \ref{table_max}.
We observe that each method upgrades a similar number of line segments, which is as expected because the budget is the same across all problems. However, the percentage reduction in risk for the MMM and CCC metrics varies between the different problem variants.

\begin{table}%
\renewcommand{\arraystretch}{1.3}
\centering
\caption{Comparing Risk Metrics.}
\label{table_max}
\begin{tabular}{llll}
\hline
& \multicolumn{3}{c}{Risk Metric Minimized} \\
& MMM & CCC & Trade-off \\
\hline
Segments Upgraded & 25 & 25 & 26 \\
\hline
MMM risk [\% reduction] & 8.76 & 8.99 & 9.43  \\
CCC risk [\% reduction] & 6.65 & 14.58 & 12.06\\
\hline
\end{tabular}
\end{table}

\begin{figure}
    \centering
    \includegraphics[width=\columnwidth]{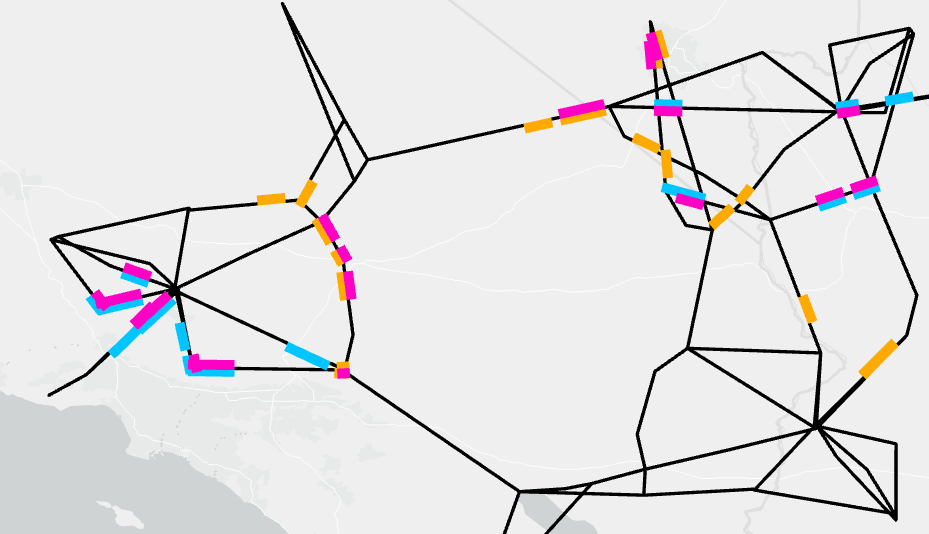}
    \caption{\small Plot of RTS-GMLC 10-km line segments selected for upgrade for the CCC (blue), MMM (orange), and trade-off (magenta) problem variants. Segments are offset for visibility.}
    \label{fig:compare_plot}
\end{figure}

To further analyze the solutions, 
Fig. \ref{fig:compare_plot} shows the 10-km line segments selected for undergrounding for the three problem variants. The sets of lines selected by the CCC and MMM formulations do not intersect, showing that the choice of risk definition has a significant impact on undergrounding decisions. Further, we observe that the trade-off variant includes a combination of line segments from each of the other solutions. This highlights that a trade-off between CCC and MMM may be useful to reduce both high point-wise and high cumulative risk. 

\subsection{Including risk-influencing factors related to power line characteristics}
As discussed earlier, the wildfire risk metrics %
can be combined with specific knowledge about the the probability of electric faults from infrastructure. We provide an illustrative example of how to incorporate the impact of voltage on ignition risk. We acknowledge that this example is simplistic (i.e. voltage is not a comprehensive measure of the relative electric fault probability), but can serve as an example of how additional (possible proprietary) grid data could be used when it is available. 

\begin{figure*}
     \centering
     \begin{subfigure}[b]{0.32\textwidth}
         \centering
         \includegraphics[width=\textwidth]{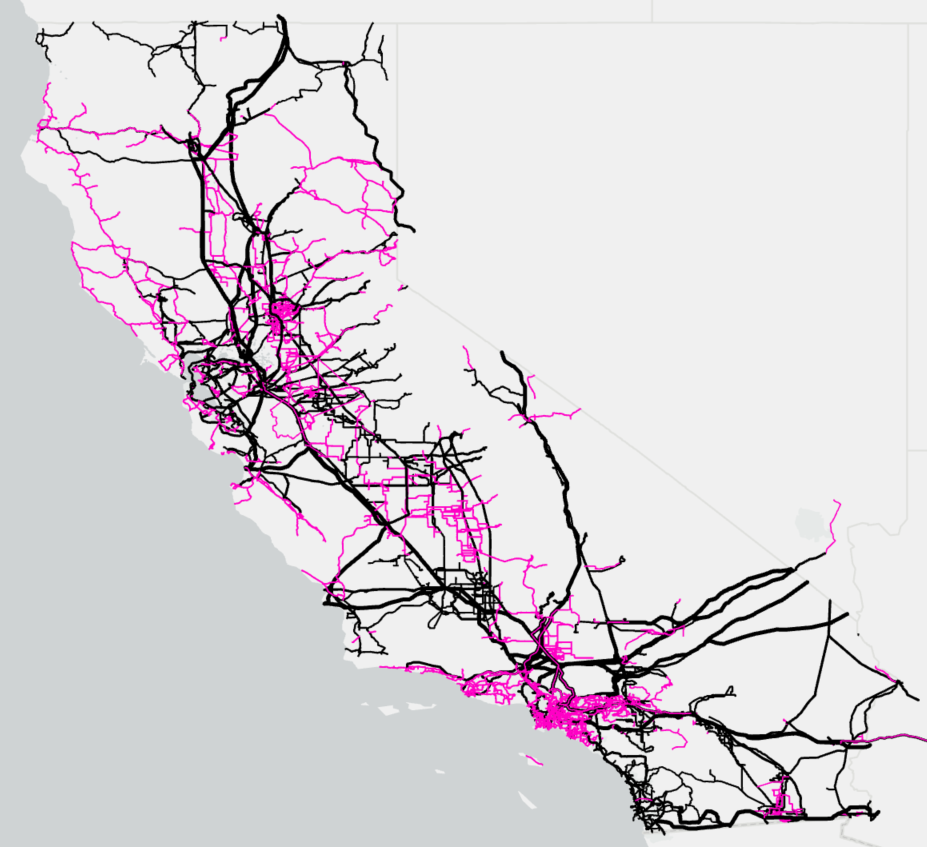}
         \caption{\small Low voltage lines (less than 69 kV)}
         \label{fig:Cal_plot_voltage}
     \end{subfigure}%
     \hfill
     \begin{subfigure}[b]{0.32\textwidth}
         \centering
         \includegraphics[width=\textwidth]{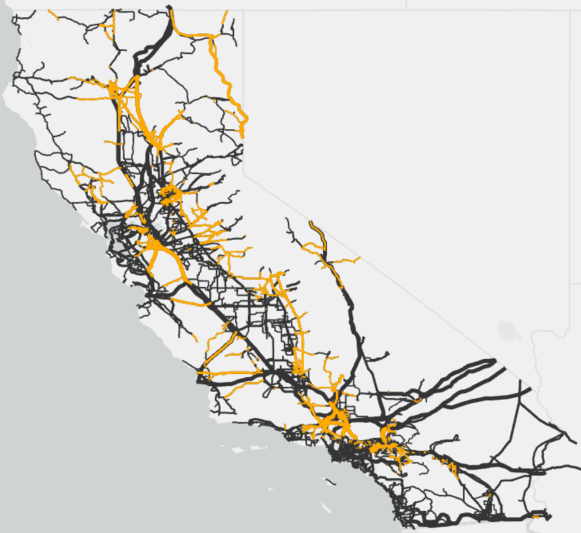}
         \caption{\small Upgrades without kV weighting}
         \label{fig:Cal_plot_without}
     \end{subfigure}%
     \hfill
     \begin{subfigure}[b]{0.32\textwidth}
         \centering
         \includegraphics[width=\textwidth]{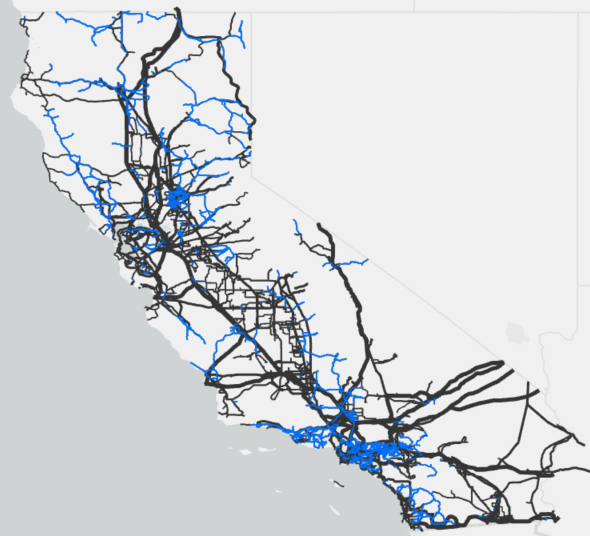}
         \caption{\small Upgrades with kV weighting}
         \label{fig:Cal_plot_with}
     \end{subfigure}
        \caption{\small Plot of California Energy Commission lines showing (a) distribution lines (less than 69 kV) in pink (b) upgrades selected without kV weighting in yellow and (c) upgrades selected with kV weighting in blue.}
        \label{fig:Cal_plots}
\end{figure*}

Specifically, we base our example on the real California grid \cite{CECdata}, which exhibit a range of line voltages from 33 to 500 kV. The system serves as a large-scale example with 6847 transmission and distribution lines. To incorporate the fact that the relative probability wildfire ignitions is three times higher for distribution grid lines, we assume that $\pi_{i,s}=1 \cdot \pi$ for transmission lines and $\pi_{i,s}=3 \cdot \pi$ for distribution lines in \eqref{eq:voltage}. 
We use $V_{\text{dist}^{\max}}=69$ kV as a cutoff value to distinguish between transmission and distribution lines.
We note that the threshold of 69kV was selected, rather than 60kV as in \cite{PGE-amendedsafetyplan}, since the CEC dataset contains very few lines with voltages lower than 60kV. In reality, it may be most accurate to use a smoother transition (i.e. multiple thresholds) for scaling that corresponds to the width of a power line's right of way. Nevertheless, this example illustrates how to incorporate these risk-influencing factors if such information is available.
Fig. \ref{fig:Cal_plot_voltage} shows the locations of the low voltage lines (less than 69 kV) in the CEC data. 

The optimization problem is solved twice for the CEC transmission system, once without any voltage-based weighting, and once with kV-weighting. The same budget of 15 billion USD is used for both cases. The lines selected for upgrade by these two solutions are shown in Fig. \ref{fig:Cal_plot_without} and Fig. \ref{fig:Cal_plot_with}.

Totals of 2536 and 3748 lines are selected for upgrade for the unweighted and kV-weighted solutions, respectively. There are 1373 lines that are exact matches, amounting to approximately 54\% of the unweighted and 37\% of the kV-weighted upgraded lines.

Based on Fig. \ref{fig:Cal_plot_without} and Fig. \ref{fig:Cal_plot_with}, we can see that the solution with kV weighting results in more lines selected for upgrade near cities, where distribution lines would be. This weighting demonstrates how we can incorporate electrical parameters such as voltage into our proposed risk assignment and upgrade selection methods. Based on the availability of risk factor data, this model could be made as specific as needed. 

\section{Conclusion}
\label{sec:conclusion}
The main contributions of this paper are a framework for assessing the wildfire ignition risk associated with power lines and an optimization model that selects lines for undergrounding such that wildfire risk is minimized. %
In particular, we discuss two different metrics for wildfire risk, and assess how the choice of risk metric impacts the optimal undergrounding decisions.
This method is simpler than gathering and synthesizing data from multiple sources to estimate the risk of ignition associated with power lines, which is advantageous when such data is complex or not publicly available. Further, the proposed method is flexible, and can incorporate grid- and component-specific information on the probability of ignition if such information is available.

One limitation of this work 
is that the WFPI indicates historical wildfire potential, not future potential. Due to climate change and human activity, historical risk may not be representative of future fire conditions. However, the proposed model is flexible enough that future wildfire risk projection maps could be used to assign risk if they were available. 

The methods and results presented in this paper are preliminary, but they provide guidance to begin to address the relatively new and challenging issue of managing wildfire ignitions from power lines. One interesting extension inspired by \cite{rhodes2020balancing} would be to model the how undergrounding reduces the need for public safety power shut-offs (i.e. allows high risk lines to remain in operation without causing excessive risk of ignitions). However, the need for public safety power shut-offs varies over time, and modeling the associated load shed would require consideration of the power flow in the system. Since the inclusion of power flow decision variables and constraints will increase computational complexity, the simpler model proposed in this paper can be used to select a subset of high-risk lines or line segments that will be considered for upgrade. %

\section*{Acknowledgement}
The authors would like to thank Lucas Franke for his help with data processing and software development. 
\bibliographystyle{IEEEtran}
\bibliography{IEEEabrv, refs}

\end{document}